\documentclass[epjST]{svjour}
\pdfoutput=1
\usepackage{graphicx}
\usepackage{floatrow}
\usepackage[caption=false]{subfig}
\usepackage{amssymb}
\usepackage{amsmath}
\usepackage[english]{babel}
\usepackage{psfrag}
\usepackage{hyperref}
\usepackage{color}

\numberwithin{equation}{section}
\begin{document}
\title{Long-Range Response to Transmission Line Disturbances in  DC Electricity Grids}
\author{Darka Labavi\'{c}\inst{} \and Raluca Suciu \inst{} \and Hildegard Meyer-Ortmanns \inst{} \and Stefan Kettemann \inst{}\thanks{\email{s.kettemann@jacobs-university.de}} }
\institute{Jacobs University Bremen, Germany
School of Engineering and Science, Jacobs University
 Bremen, Bremen 28759, Germany}
\abstract{
We consider a DC electricity grid composed of transmission lines connecting power generators and consumers at its nodes. The DC grid is described by nonlinear equations  derived from Kirchhoff's law.  For an initial  distribution of consumed and generated  power, and given  transmission line conductances,
we determine the geographical distribution of  voltages at the nodes.  
 Adjusting the  generated power for the Joule heating losses, we then 
  calculate  the electrical power flow through the transmission lines.  Next, we study the response of the grid to  an additional transmission line between two sites of the grid and calculate the resulting  change in the power flow distribution.  This  change  is found to decay slowly in space, with a power of the distance from the additional line. 
We find  the geographical distribution of the  power transmission, when a link is added.  With  a finite probability  the maximal load in the grid becomes larger when 
  a transmission line is added, a phenomenon that is known as  Braess' paradox. 
   We find that this phenomenon is more pronounced in  a DC grid described by the 
    nonlinear equations  derived from Kirchhoff's law  than in a linearised
     flow model studied previously in Ref. \cite{witthaut2013}.
 We observe furthermore that  the increase in the load of the transmission lines due to an added line is of the same order of magnitude as  Joule heating. Interestingly, for a fixed system size the load of the lines increases with the degree of disorder in the geographical distribution of consumers and producers.
}
%
\maketitle
\section{Introduction}
The stability of electricity grids requires to protect it against fluctuations of the electricity generators and consumers, and disturbances in the transmission lines \cite{kundur,amin}. Therefore, the electrical power system must be constructed in such a way that,  when subjected to a physical disturbance,  it regains an operating equilibrium without exceeding bounds in the fluctuations of the system variables. Since this is a highly complex and nonlinear problem, the study of its dependence on the network topology, the operating conditions and forms of disturbances requires to make modeling assumptions \cite{kundur}. Recently, the Braess' paradox, known from traffic flows, has been found to be relevant in power grids as well \cite{timme}. In the context of power grids the Braess' paradox amounts to a decrease in the overall performance, although a transmission line was added rather than removed. The reason is that the added line may lead to an increase of  the load in some other transmission lines, even beyond their maximal capacity. On the other hand, it was found that the danger of a blackout, the total disruption of the electricity grid, can be reduced by decentralisation of the power generation \cite{timme,witthaut}.

A realistic model of electricity grids should take into account the voltage fluctuations as well as fluctuations in the incoming and outgoing electrical power \cite{kundur}.
 In AC grids random phase fluctuations of the impedances  and frequency must be considered as well \cite{kundur}. As a first step towards a prediction of the stability of realistic power grids against a change in the transmission lines  we study here DC power grids and  study 
the power flow through all links of the network as described by a set of nonlinear equations that are equivalent to  Kirchhoff's law at each site. We then add a single transmission line in the center of our regular grid and analyze the spatial dependence of the induced change in the load of the transmission lines.

 Thereby,  we aim  to address  the following questions about the stability of the network:  how is the power  transmitted through the transmission lines distributed and what is the probability to 
  come close to its power capacity limit as a function of the 
   network parameters and the distribution of the  consumer and generator power?
   How does that distribution   change, when  one transmission line  is added? In particular, we are interested in the spatial distribution of the resulting changes in transmission power, and  how this change decays with the spatial  distance $r$ to the perturbation. Does it  typically decay exponentially or with a power law with that distance? A power law decay would indicate a nonlocal effect of the perturbation on the power grid stability.

Towards this end our strategy  is the following: we choose a realistic value $V_0$ for the nominal voltage, 
 which is the voltage to be received by the consumers, and the conductance of the transmission lines. 
For a given randomly chosen distribution of consumers and generators $\{P_i^0\}$ and the given value $V_0$ we  determine the corresponding set of voltages $\{V_i\}$ from  a linearised set of equations. We next choose the resulting set of voltages $V_i$ as starting point to determine the induced power flow $F_{ij}$ through the network
 links $(i,j)$,   including  the   Joule heat  $dP_{ij}^\Omega$. This flow is the quantity of our main interest, in particular how this flow changes if one transmission line is added. Along with that, we obtain  the additional power $dP_{i}$ that has to be produced at a site $i$ to compensate for the loss via Joule heat.  When the  generator power is adjusted that way,  the set of voltages $\{V_i\}$  then solves   the  nonlinear set of equations for $V_i$, which follow from Kirchhoff’s laws, and we can analyse the  results for  the power transmission $F_{ij}$.

\subsection{DC Power Grid}
Let us consider a DC power grid with N sites denoted by the index $i$. The conservation of power yields at every node the equation
	\begin{equation}
            P_i = \sum_{j \neq i} F_{ij}
	\end{equation}
for all nodes $i =1,..,N$. Here, $P_i>0$, if there is an electricity generator at site $i$, while $P_i < 0$, when power is consumed at site $i$. $F_{ij}$ is the power transported from site $i$ to site $j$. When the voltage at site $i$ is $V_i$, the transmitted power $F_{ij}$ is related to the electrical current $I_{ij}$ between sites $i$ and $j$ by $F_{ij} = V_i I_{ij}$. These currents are related to the voltage difference by Ohm's law $I_{ij}/G_{ij} = V_i - V_j$, where $G_{ij}$ is the conductance of the transmission line between sites $i$ and $j$. Thus,  choosing the local power $P_i$ for all sites $i$ and the conductances $G_{ij}$, the voltages $V_i$ are determined by the equations
	\begin{equation}  \label{dc}
              P_i = V_i \sum_j G_{ij} (V_i- V_j),
    \end{equation}
which  are N nonlinear equations for the N voltages $V_i$. We can rewrite these equations, by relating the power $P_i$ to currents $I_i$, which are incoming/outgoing at sites $i$ as $P_i = I_i V_i$. Inserting this in Eq. (\ref{dc}), we obtain
	\begin{equation}  \label{idc}
                  I_i =  \sum_j G_{ij} (V_i- V_j),
	\end{equation}
which is nothing than Kirchhoff's law at site $i$. If we consider for a given electricity grid the  distribution of incoming and outgoing power $\{P_i\}$ as given (rather than the currents $I_i$),  we need to solve Eq.  (\ref{dc}), which is nonlinear in $V_i$. The power loss due to Joule heating in transmission line $(i,j)$ is given by
	\begin{equation}\label{eqohm}
		 dP_{ij}^\Omega= G_{ij}(V_i-V_j)^2=  F_{ij}+F_{ji}.
	\end{equation}
Here, the link $(i,j)$ is oriented such that $F_{ij} >0$. It is this  Joule heating at link $(i,j)$ that should be compensated for by the power generators. In realistic transmission lines  the Joule heating $dP_{ij}^\Omega$ does not exceed several percent of the transmitted power  $F_{ij}$ under stable operation conditions.

It should be noticed that the authors of \cite{witthaut2013},  who demonstrated the Braess'paradox in a flow model, consider a different set of equations (Eq. 20 of Ref. \cite{witthaut2013} in  Appendix A2),  which is linear and corresponds in our notation to
	\begin{equation}\label{eq14}
		P_i^0\equiv V_0 \sum_j G_{ij}(V_i-V_j),
	\end{equation}
with  $\sum_i P_i^0=0$. Here $V_0$ is the nominal grid voltage.  This equation was  derived from a Lagrangian by minimizing the total dissipative power under the constraint of energy conservation at site $i$, $\sum_{j=1}^N F_{ij}=P_i$, with $F_{ij}$ the power transmitted from node $i$ to node $j$. We note that the physical Kirchhoff's laws for DC electricity grids rather result in the nonlinear equations Eq. (\ref{dc}). Our motivation to reconsider the problem studied in Ref. \cite{witthaut}  was to compare the order of magnitude of the Joule heating with the  changes in the transmitted   power due to an added link capacity, in particular to see whether the inclusion of the Joule heating increases or decreases the chance for a Braess paradox to occur in the DC grid. So we choose the quadratic equations Eq. (\ref{dc}) as our starting point.

\vskip5pt In order to find  solutions to these nonlinear equations  we proceed along  the following steps:
  \begin{itemize}
  \item We first solve the linearised equations Eqs. (\ref{eq14}) for an ideal DC grid with $\sum_i P^0_i =0$, to find the  set of  $V_i$ at  all nodes $i$ for a randomly chosen distribution $\{P_i^0\}$.
  \item Next we use this set of  $V_i$ to  calculate the total  power transmitted from node $i$ into the link $(i,j)$     as given by
	\begin{equation}\label{eq:nonlF}
		F_{ij}= V_i G_{ij} \left( V_i - V_j \right),
	\end{equation}
where  the power flows  from i to j, when $V_i > V_j$.
  \item   In order to solve the nonlinear equations Eq.~(\ref{dc}) by this set of $\{V_i\}$ the power distribution $\{P_i^0\}$ must be modified  to an  adjusted set of $P_i$, by adding 
  \begin{equation}
  d P_i (\{V_j\}) = (V_i - V_0)  \sum_{j} G_{ij} \left( V_i - V_j \right),
  \end{equation}
    on each node, so that the adjusted $P_i$ in Eq.~(\ref{dc}) results from the given $\{P_i^0\}$ and the calculated $\{dP_i\}$. Summing over all nodes, we find $\sum_i P_i = \sum_i dP_i=\sum_i dP_i^{\Omega} = dP^{\Omega}$, where $dP_i^{\Omega}$ is given by  the sum over $j$ of Eq.~(\ref{eqohm}). (Note that according to this definition  the double sum in $\sum_{i} dP_i=\sum_{i,j}F_{ij}$ is unrestricted in $i$ and $j$, while the double sum in $\sum_i dP_i^{\Omega}=\sum_{ij}(F_{ij}+F_{ji})$ runs only over directed links $(i,j)$,depending on the relative size of the voltages.) Thus, as expected, $dP^{\Omega}$   is the  total power dissipated as Joule heating in the electricity grid, which  must be additionally produced by the power engines if one wants to guarantee that the consumers get the needed power.
  \end{itemize}

Note that the following conditions must be imposed for stable grid operation:
  \begin{enumerate}
  \item Joule heating $d P_{ij}^{\Omega}$ must be smaller than the power injected in link $(i,j)$, $F_{ij}$. This yields              $V_i-V_j < V_i$,     or $V_j > 0$ for all $j$.
  \item  Joule heating $d P_{ij}^{\Omega}$ should not exceed the power capacity, here chosen as $F_{ij}^{max} = V_0^2 G_{ij}$ (as if the maximal voltage drop off over a line is determined by the nominal voltage). This gives $|V_i-V_j| < V_0$ for all $(i,j)$.
  \item  The total power capacity of all transmission lines connected to node $i$ should exceed the injected power $P_i$,                yielding $ V_0^2 \sum_j G_{ij}  \gg  |P_i|$.

  \item Breakdown of a transmission line occurs if the transmitted power $F_{ij}$ approaches or  exceeds the                power capacity $F_{ij}^{max}$. Thus, to ensure grid stability one needs to impose $F_{ij}< F_{ij}^{max}$. This gives the condition $|V_i (V_i- V_j)| < V_0^2$ for all $(i,j)$.
  \end{enumerate}
With the equations Eq. (\ref{dc}) we can answer the following questions about the stability of the network: How close is the power transmitted via the link $(i,j)$, $F_{ij} = G_{ij}V_i  (V_i -V_j)$ to the power capacity  $F_{ij}^{max}=V_0^2 G_{ij}$ of that link? In order to study this systematically, we then find the distribution of $|F_{ij}|$. How does the distribution of $F_{ij}$ change, when a link $(m,n)$ with conductance $G_{mn}$ is added? In particular, we are interested in the spatial distribution of the resulting changes $\Delta F_{ij}$, and in how this change decays with the spatial distance $r$ to the perturbation. Does it  typically decay exponentially or with a power law with that distance? A power law decay would indicate a nonlocal effect of the perturbation on the power grid stability.

\begin{figure}[t]
\captionsetup[subfigure]{labelformat=simple}
\centering
\subfloat[\label{sfig:radial_generala}]{\includegraphics*[width=0.238\textwidth]{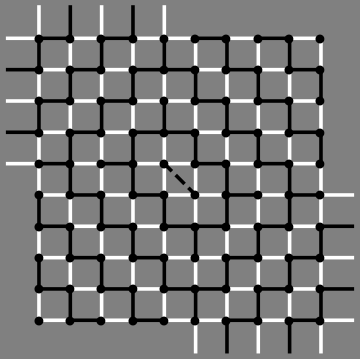} }\hfill
\subfloat[\label{sfig:radial_generalb}]{\includegraphics*[width=0.3\textwidth]{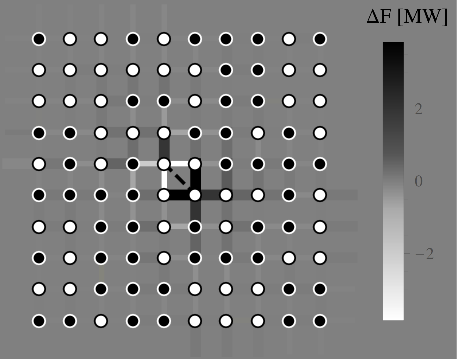}}\hfill
\subfloat[\label{sfig:radial_generalc}]{\includegraphics*[width=0.3\textwidth]{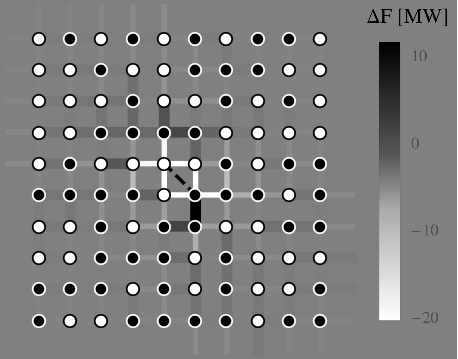}}\hfill
\caption{ a) Geometry of a $10 \times 10$ lattice, where black/white links are equidistant  by an
even/odd number of links $r$ from the
  added link (dashed line in the centre of the grid).
 b)~Distribution of the change  in the  power transmission $\Delta F_{ij}$, after adding a link between two  consumers  (white dots) as indicated by the  lines whose  white/gray intensity corresponds to the change in MW as defined in the color bar.
 c)  Distribution of the change  in the load after adding a link between a consumer (white dot) and a generator (black dot).} \label{fig:radial_general}
\end{figure}

\begin{figure}[t]
\centering
\includegraphics*[width=0.5\textwidth]{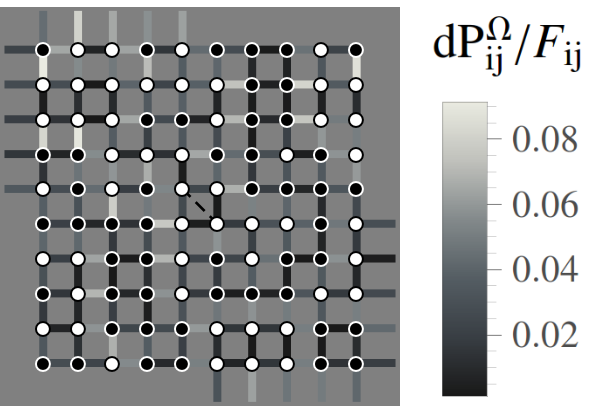}
\caption{Distribution of the ratio of  Joule heating  $dP^{\Omega}_{ij}$ and the transmitted power $F_{ij}$. Light/dark color represents a higher/lower ratio for the unperturbed grid.} \label{fig:Ohm_over_load}
\end{figure}

\section{Numerical Results} \label{sec:results}

We choose a square lattice with dimensions $L_x \times L_y$ with periodic boundary conditions, shown in Fig.~\ref{sfig:radial_generala}, with a random geographical distribution of producers ($P^0_{p}=P_0$, black dots) and consumers ($P^0_{c}=-P_0$, white dots), which satisfies the condition $\sum_i P_i=0$, $i=1,...,N$. Each node is connected to the four nearest neighbours by transmission lines with conductance $G_{ij}$, which we take as $G_{ij} = G_0 A_{ij}$, where $A_{ij}$ is the adjacency matrix of the square lattice.  We choose $G_0$  such that the losses due to the Joule heating in  the transmission lines are of the order of $1\%$ of the power $P_0$. In order to calculate the load $F_{ij}$ of each link, we solve the system of $N$ linear equations~\eqref{eq14}, where $N$ is the total number of nodes, $N = L_x L_y$. The rank of the system \eqref{eq14} is $N-1$, so a solution, if it exists, has one of the $V_i$ undetermined. For every configuration we choose the set of voltages $\{V_i\}$ such that  the minimal voltage $min\{V_i\}=V_0$ is guaranteed, so that the consumer gets at least the nominal voltage.

  In order to study how the change in the load after adding an additional link spreads on the lattice, we measure the change in the load $\Delta F_{ij} = F_{ij}^{after} - F_{ij}^{before}$ as a function of the distance from the added link. Let $l_m$ and $l_n$ represent nodes on the lattice between which the link is added. We define the radial distance $r$ of a link from the added link as the minimal number of steps required to reach that link from  one of the nodes $l_m$ or $l_n$.
	\begin{figure}[h]
	\includegraphics*[height=5cm]{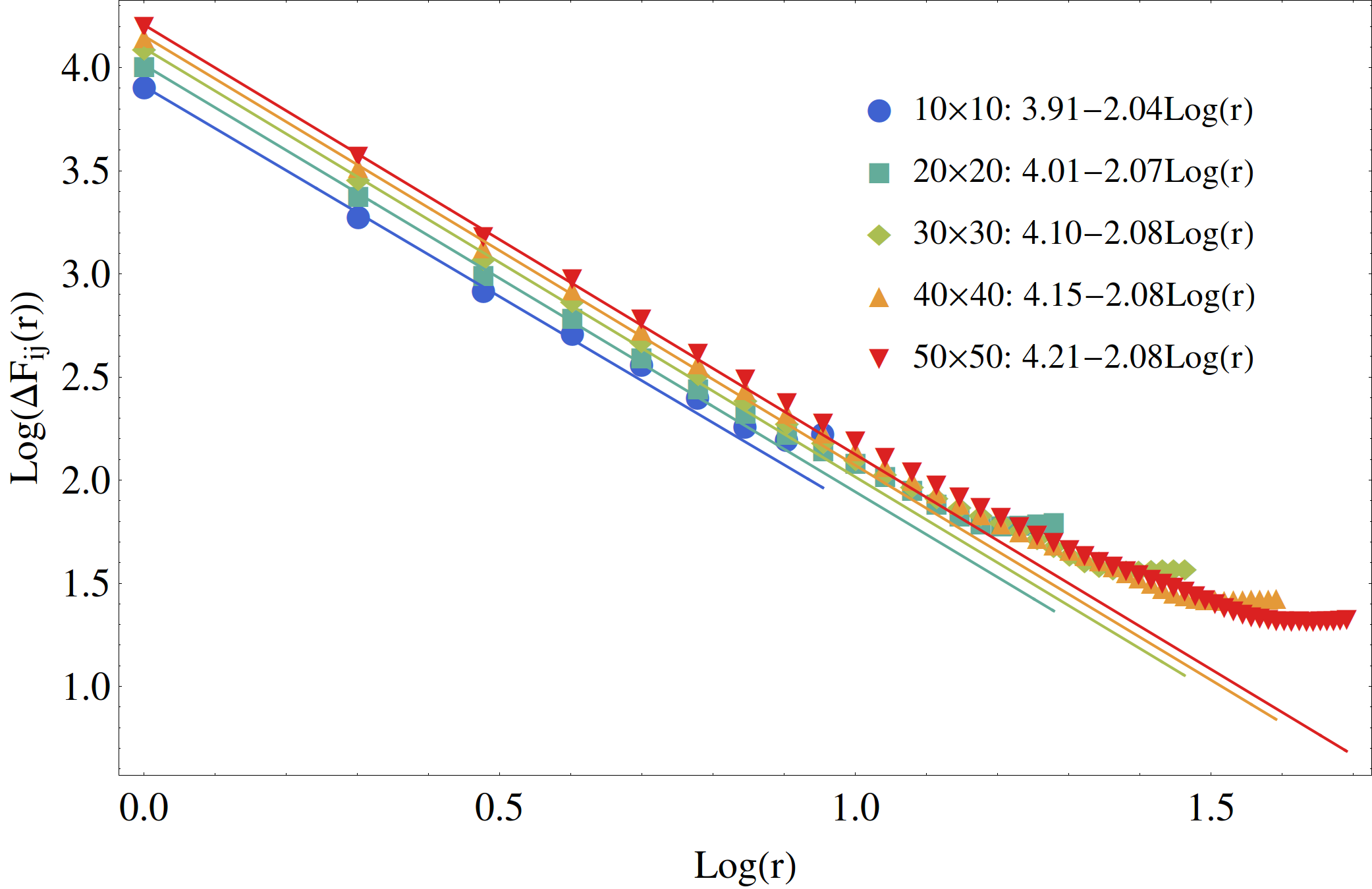}
	\caption{Color online. Average of the power flow $\langle |\Delta F_{ij} \left (r\right)| \rangle$ as function of the distance $r$ to the added
	 transmission line  for different sizes $L \times L$
	 with,  from bottom to top, $L=10, 20, 30, 40, 50$, respectively.  Numerical results (points) are shown together with the  fit to
	  $f(r) = ar^{-b}$ (values for $a$ and $b$ are given in the inset according to $\ln a-b\log r$).} \label{fig:radial}
	\end{figure}
Equidistant  links around the added link (dashed line)  whose distance $r$ is an even/odd number are plotted by black/white
lines  in Fig.~\ref{sfig:radial_generala}  for a  $10 \times 10$ square lattice with periodic boundary conditions. Fig.~\ref{sfig:radial_generalb} and \ref{sfig:radial_generalc} show  typical distributions of $\Delta  F_{ij}$ on a $10 \times 10$ lattice. There is a difference in the distribution of $\Delta F_{ij}$, when we add a link between two nodes of different type (producer and consumer), Fig.~\ref{sfig:radial_generalc}, and between the same types (two producers, or two consumers), Fig.~\ref{sfig:radial_generalb}. Typically an added link between nodes of different type is surrounded by links whose load has decreased, Fig.~\ref{sfig:radial_generalc}, while an added link between nodes of  the same type is surrounded by both links with  decreased and increased transmission power. 

We fix the parameters to  $V_0 = 10 kV$, $P_i= \pm 100 MW$. The conductance  $G_0 = 10/\Omega$ is chosen to satisfy the condition that the loss due to Joule heating is less than $10\%$ of the power transmission per link. With this choice of parameters the voltage differences are  $\Delta V_{ij} = V_i-V_j  < 1 kV$, so that the Joule heating of the link $(i,j)$, $d P^{\Omega}_{ij}= G_0\Delta V_{ij}^2 < 10 ~ kV^2/\Omega  = 10 MW = 10 \%$ of $P^0$. Fig.~\ref{fig:Ohm_over_load} shows the distribution of Joule heating on a lattice relative to  the load $F_{ij}$. We find that for these parameters, it does indeed  not exceed about $10 \%$ of the transmitted power.

\begin{figure}[h]
\center
\includegraphics*[height=4.5cm]{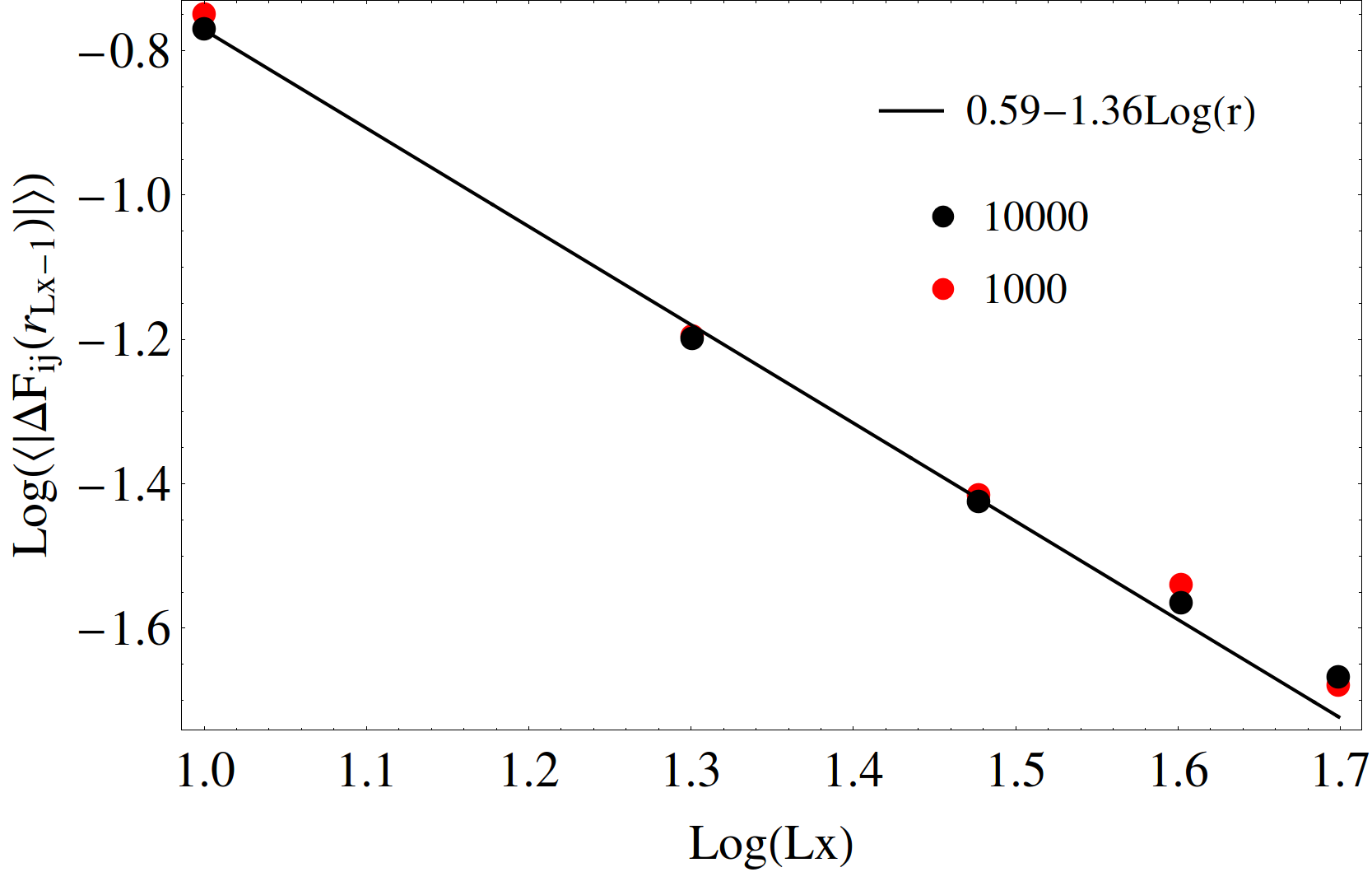}
\caption{Color online. Saturation value  $\mid \Delta F_{ij} (r=L-1) \mid$ as a function of $L$. } \label{fig:saturation}
\end{figure}

\begin{figure}[h]
\includegraphics*[height=5cm]{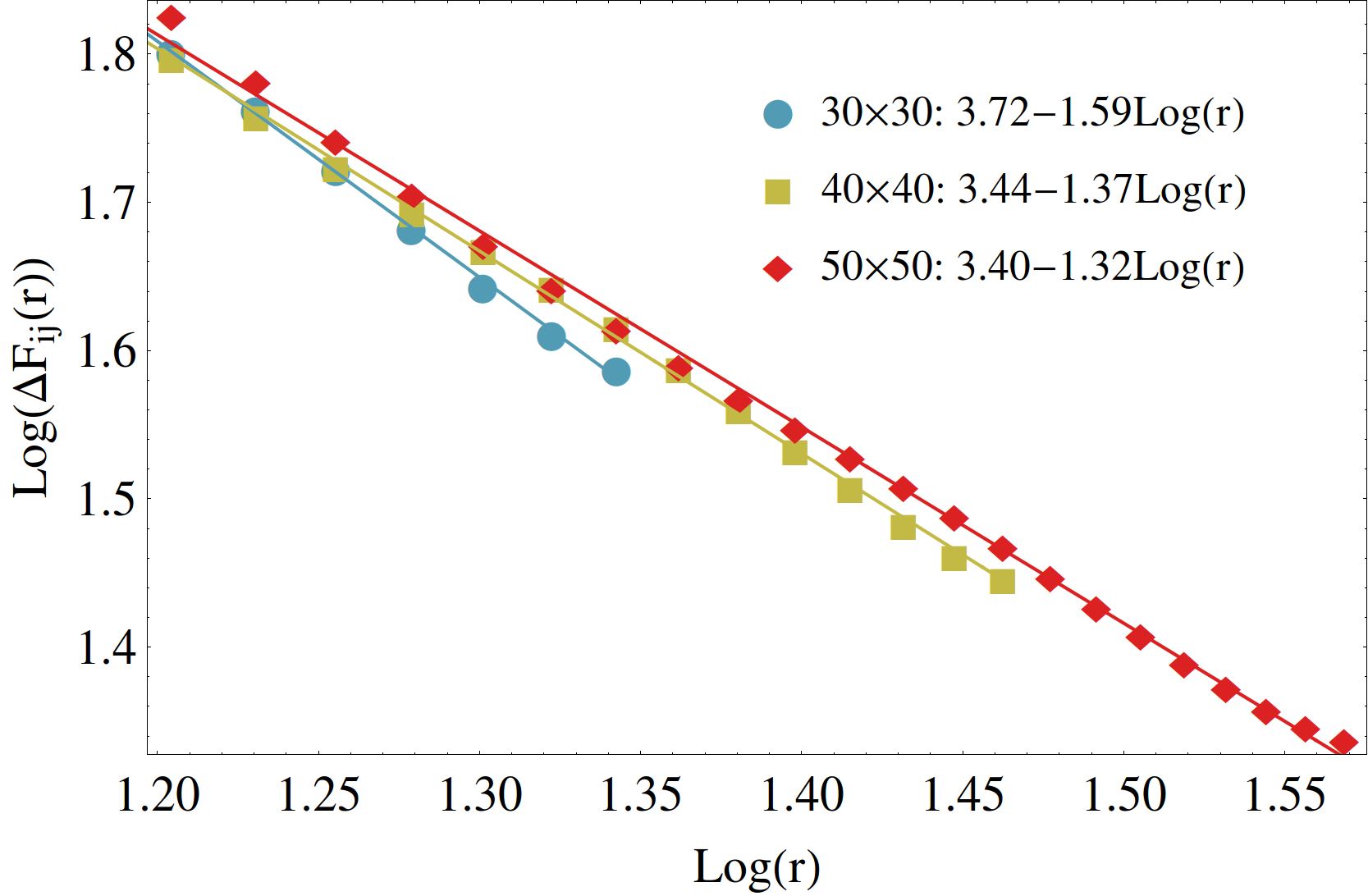}
\caption{Color online. Zoom into the intermediate distance regime,  where data for  $\mid \Delta F_{ij} \left (r\right) \mid$ obey a power law $f(r)=ar^{-b}$, with a smaller power, see the inset (data  in  the saturation regime are here not shown).} \label{fig:radial_mid}
\end{figure}

\begin{figure}[ht]
\captionsetup[subfigure]{labelformat=simple}
\center
\subfloat[\label{sfig:first_rada}]{\includegraphics*[height=4.cm]{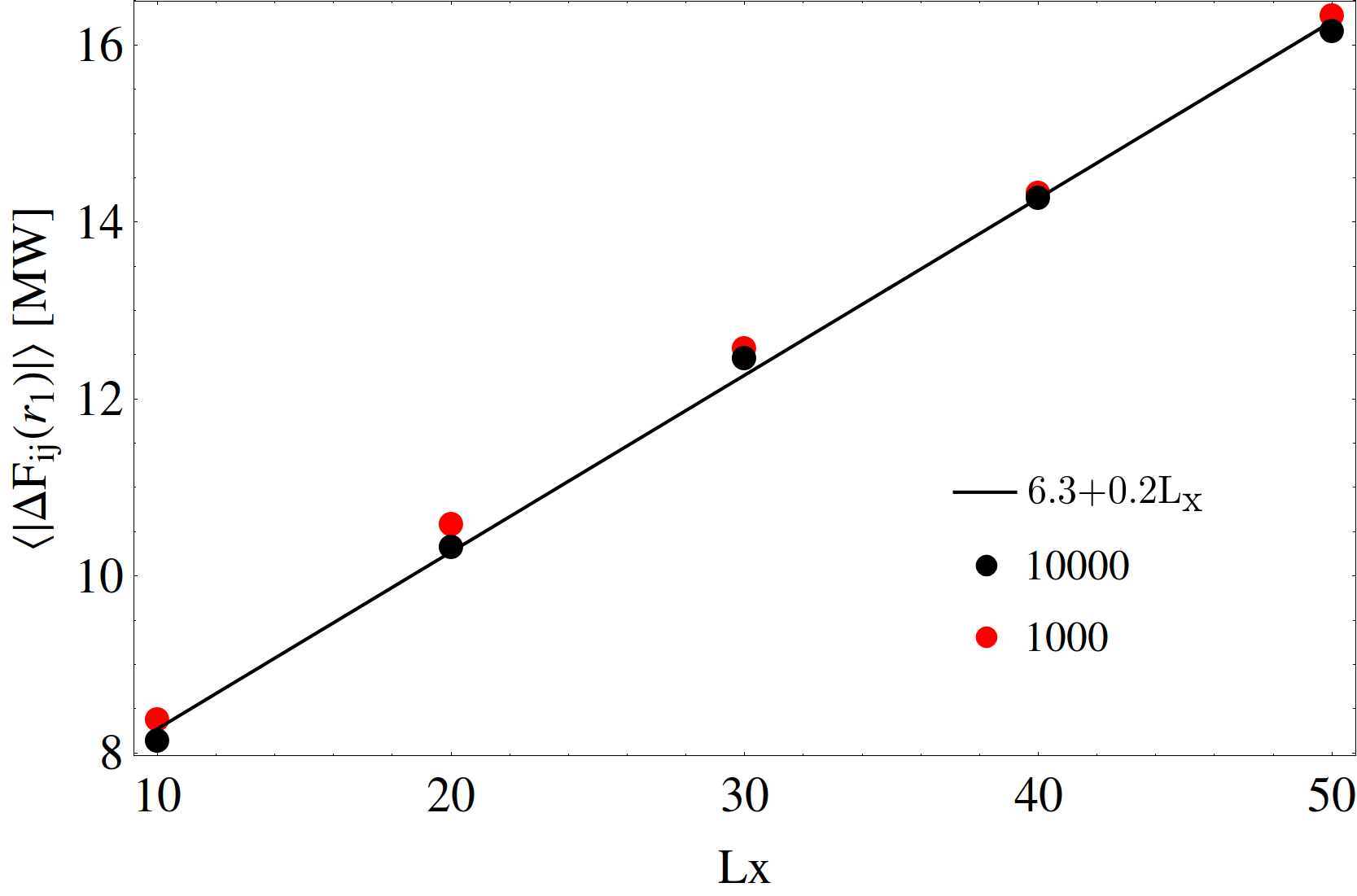}}\hfill
\subfloat[\label{sfig:first_radb}]{\includegraphics*[height=4.cm]{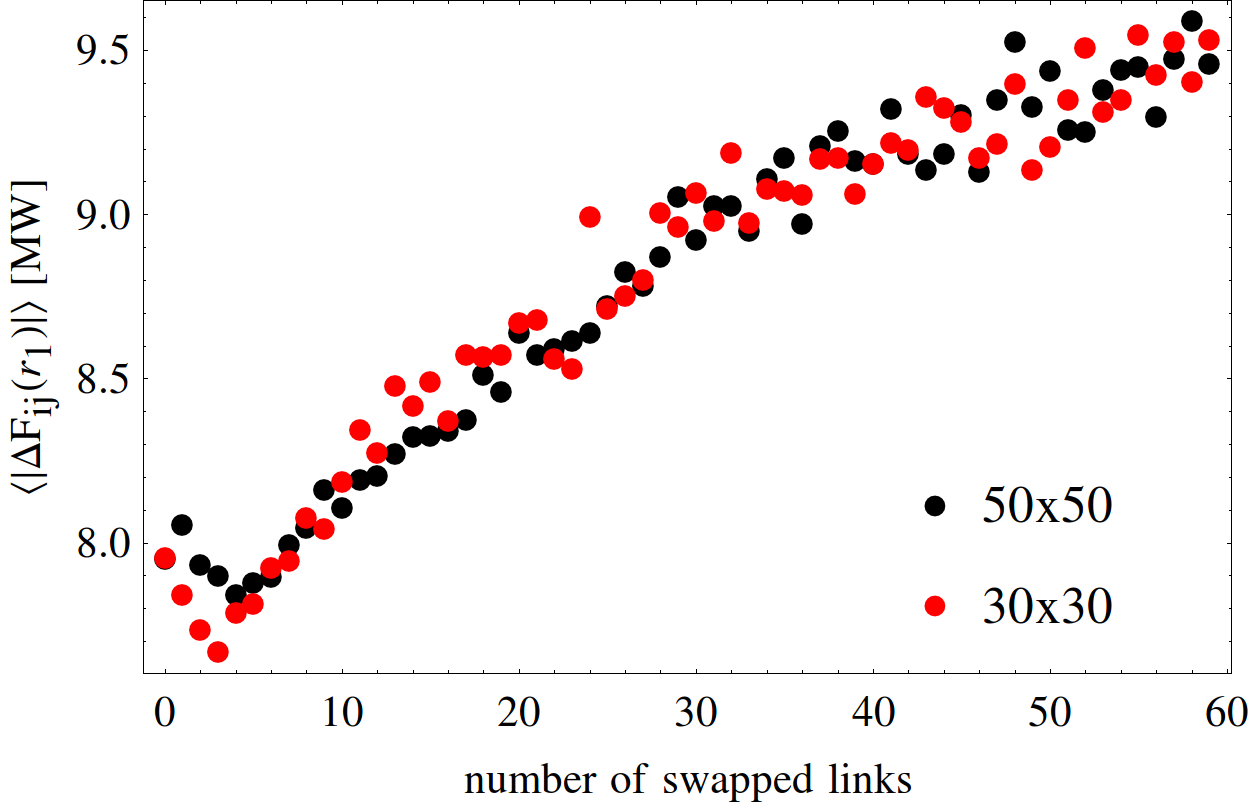}}\hfill
\caption{ Color online.
a) Change in the load at $r=1$ as a function of the system size $L$. Black dots represent data averaged over 10000 realizations of the power distribution, red (grey) ones over 1000. b) Change in the load at $r=1$ as a
 function of the number of swapped links  for $L=10$ to illustrate the effect of disorder.
 } \label{fig:first_rad}
\end{figure}

\begin{figure}[h]
\center\captionsetup[subfigure]{labelformat=simple}
\center
\subfloat[\label{sfig:histograma}]{\includegraphics*[height=5.2cm]{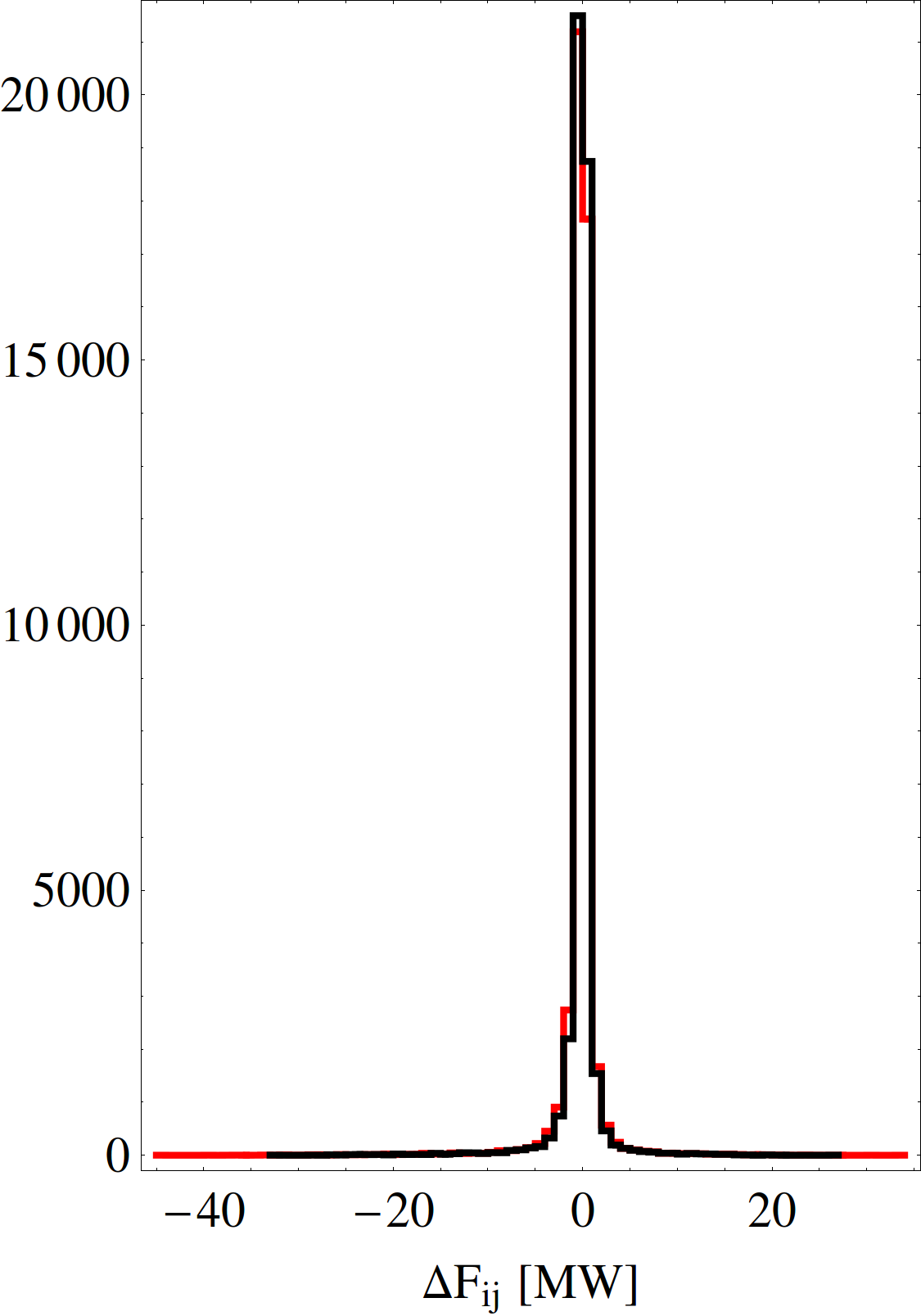}}\hfill
\subfloat[\label{sfig:histogramb}]{\includegraphics*[height=5.2cm]{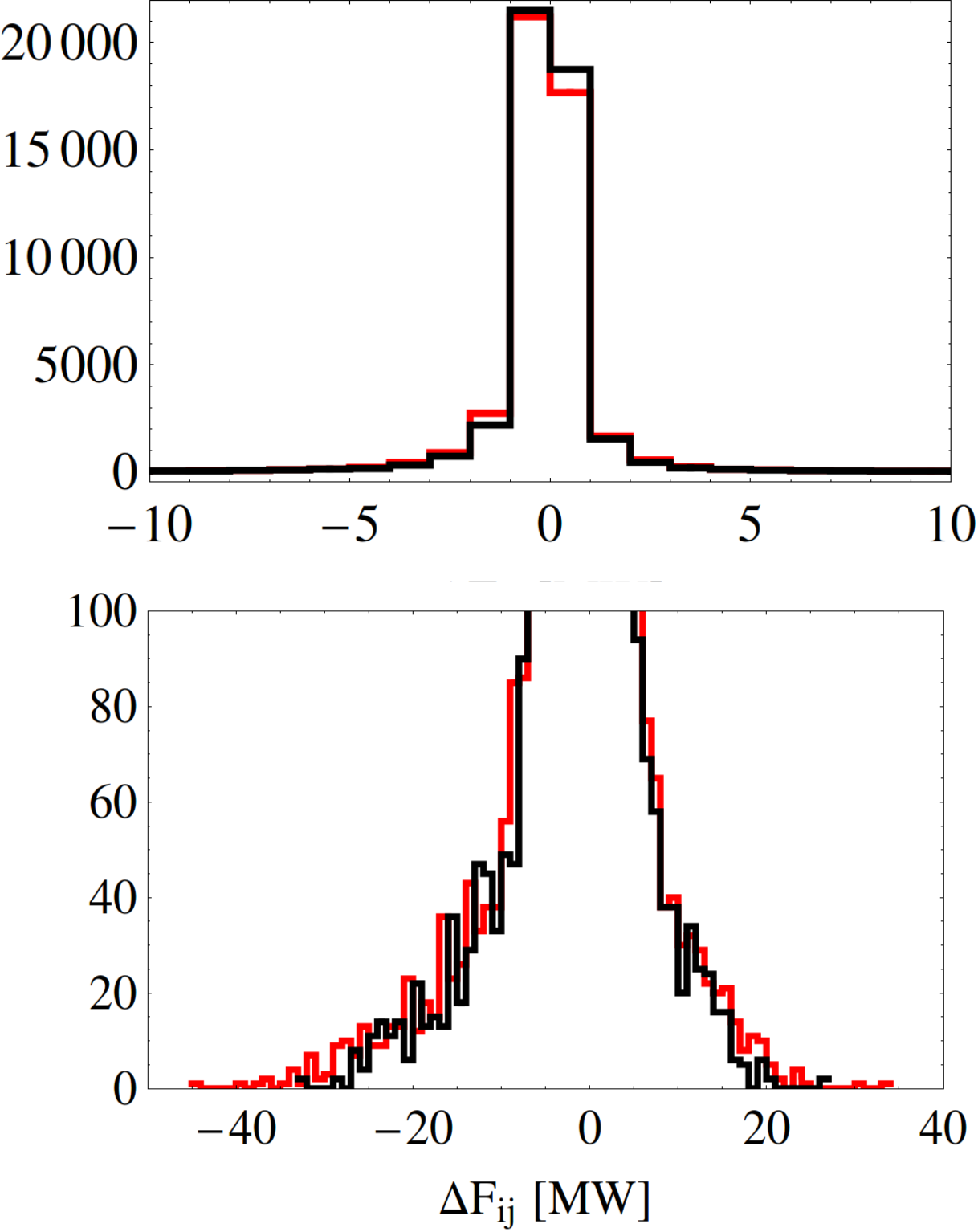}}\hfill
\subfloat[\label{sfig:histogramc}]{\includegraphics*[height=5.2cm]{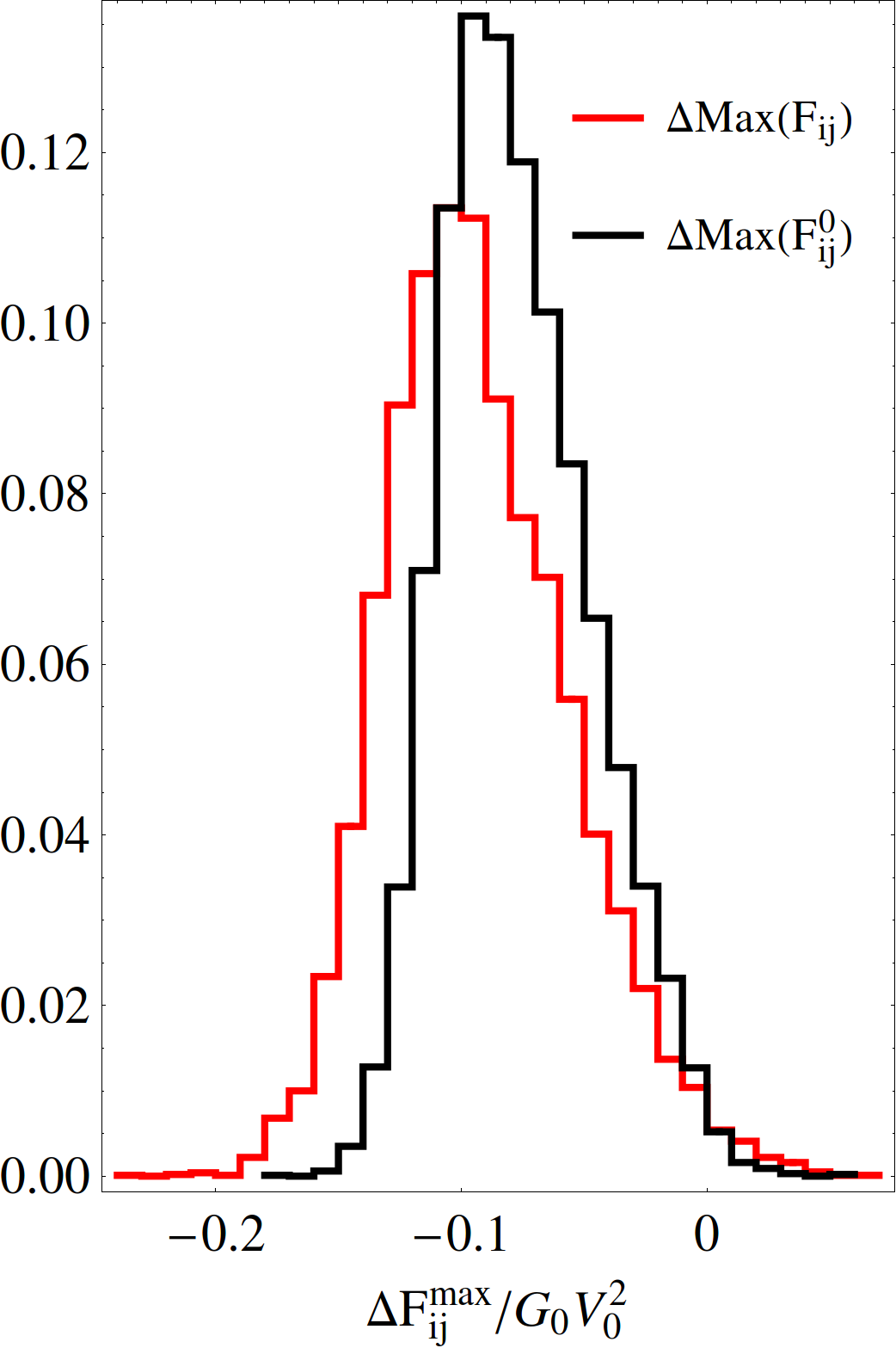}}\hfill
\caption{Color online. a) Histogram of the change of $F_{ij}$ on all links
 after adding a link  for  10000 realizations of ${P_i^0}$ in a 10 $\times$ 10 grid.
 Positive (negative) values represent an increase (decrease) of the load.  b) Zoom around the peak value (upper figure) and tails of the distribution (lower figure). c) Probability  of the change in the maximum  load after adding a link, relative to  the power capacity $V_0^2G_0$.
 A finite probability for a positive value indicates Braess' paradoxon. 
 }\label{fig:hist_dF}
\end{figure}

In order to get the average power transmission change as a function of the distance $r$, we next sum the absolute values of all changes  $|\Delta F_{ij} (r)|$ at links at the  same distance $r$, divide it by the number of such  links $M_r$, and average over 10000 realizations of the power distribution $\{P_i\}$. Fig.~\ref{fig:radial}~ shows the resulting average  change $\langle |\Delta F_{ij} (r)| \rangle$ as a function of the distance $r$. The data are fitted with a power function
$f(r) = ar^{-b}$,  in a double logarithmic plot. We can clearly distinguish different short and long range behavior. For small $r$, at a distance of a few lattice constants, we find a power law behaviour with power $b \approx 2.1$, see the inset of Fig.~\ref{fig:radial}.

We find that for a given system size $L$, $|\Delta F_{ij}|$ saturates at large distances $r\rightarrow L$. The saturation here refers to the fact that it no longer decays but fluctuates around a certain value over a few sites before the maximal distance is reached. However, this  saturation value of  $|\Delta F_{ij}|$  depends on the system size and decays with $L$, as shown in Fig.~\ref{fig:saturation}, where $|\Delta F_{ij} (r_{\rm max})|$ is plotted  taking  the value at the  largest possible  distance to the perturbation $r_{\rm max}$ as a function of $L$.  This clearly  confirms  that this saturation is a finite size effect.

Plotting only the values $\langle|\Delta F_{ij}(r)|\rangle$ at intermediate distances $r$, disregarding the data in the saturation region in order to avoid this finite size effect, as done in Fig.~\ref{fig:radial_mid}, we find that  all  data for different system sizes $L$ fit  again a power law $f(r)=ar^{-b}$, but with a smaller power $b$ approaching $ b \approx 1.3$ 
 for the largest grids considered here.

  Next, we study another effect which depends on the systems size $L$: 
Fig.~\ref{sfig:first_rada} shows $|\Delta F_{ij}|$ for the smallest radius $r_1=1$ as a function of $L$ as  averaged over 1000 and 10000 ensembles of power distributions. We notice a systematic linear  increase of the $|\Delta F_{ij} \left (r_1\right)|$ with the  increase of $L$. This 
 increase is  due to the  increase in the voltage difference as the  grid size increases. It is challenging to trace back, in which way the larger system size leads to an average increase in $V_i-V_j$ between neighboured sites close to the perturbation, although the equations do not show an obvious source of  size dependence. In view of that we look at the voltage distribution for a fixed lattice size, but consider different power distributions,
  which  differ by their degree of order. We start with an ordered $L\times L$ lattice, for which entire rows of  consumers alternate with rows composed of only producers. We then choose randomly between one and $L$ (producer, consumer)-pairs and switch their positions to induce more and more disorder
  in the geographical distribution of consumers and producers. In Fig.~\ref{sfig:first_radb} we plot the  maximum value of voltages for $L=30$ and $L=50$  as a function of that kind of power distribution  disorder. As the minimum voltage  is fixed to $V_0$, an increase of the maximum voltage  ${V_i^{max}}$ corresponds to  an increase of the maximum difference of the voltages, which is proportional to $max\{F_{ij}\}$. So for a fixed system size and different degrees of randomness in the power distribution, we see a tendency of an increase of $max\{F_{ij}\}$ with increasing randomness. On the other hand, if we compare two systems with the same kind of random distributions of $\{P_i\}$, which just differ by their size, our conjecture is that one of the reasons for an increase in the load around the perturbation is the increased total amount of disorder in a larger system.  This conjecture will be further pursued in a future publication. Another reason 
   for such an increase of the load in the transmission lines  can be  due to an increasing resistance of the electricity grid with the system size, and the resulting increase of the maximal voltage ${V_i^{max}}$ when the minimal voltage is fixed to $V_0$.\\

  Next, we study the distribution in  the change in transmitted power when a transmission line is added. In particular, we are interested in the probability with which 
   Braess paradox occurs, meaning  that the transmission  line  which  transmits the 
   largest amount of   power  transmits even  more power after amother  line was added. 
In Fig.~\ref{sfig:histograma} we show the distribution of the change in the transmission power  $\Delta F_{ij}$ in units of MW  as a histogram obtained from all the transmission lines $(i,j)$ of the electricity grid, except for the added one. For comparison, we also show the histogram of the change in $\Delta F^0_{ij}$,  the change of  $F^0_{ij} = V_0 G_{ij} (V_i-V_j)$ (which  is the quantity studied in Ref. \cite{witthaut}). We note that the distribution of
 the   change in the real power transmission  $\Delta F_{ij}$, including Joule heating, has a wider distribution with longer tails to positive, and more prominently to negative values. In Fig.~\ref{sfig:histogramb} we show zooms around the peak value (upper figure) and the tails of the distribution (lower figure) to  show this small effect  more clearly. In Fig.~\ref{sfig:histogramc} we plot the  probability  for  a  change 
  of the maximal transmitted power  $\Delta {\rm Max}(F_{ij})$ 
  relative to  the power capacity $V_0^2G_0$, as obtained from 10000 ensembles.
 A finite probability to have a positive change in the transmitted power 
  indicates Braess' paradoxon.   
     For the chosen parameter values, the load still remains within $25\%$ of the power capacity limit of the transmission lines, so that the overall performance is not seriously affected.  Choosing  different initial parameters, in particular,   increasing the injected power  $P_0$, would 
        bring the maximal power transmission closer to the transmission capacity and the addition of a line can result in a  power outage of the electricity grid. 

\section{Conclusions} \label{sec:conclusions}
We  studied  the response of a  square lattice grid to  an additional transmission line between two sites of the grid. We  calculated the induced change in the power flow distribution, and found that it  decays slowly,  with a power of the distance from the additional line.  The power law exponent  at small distances, $b= b_n \approx 2.1$,
 is larger   compared to  the one obtained at long distances, where we find $b= b_l < 1.6$, approaching 
 $ b_l \approx 1.3$ in the  largest  grid. We therefore  conclude   that the  addition of  a  link has a long-range effect, at least  on the   square electricity grid model  with nearest neigbour coupling studied here. 
When the spatial  distance $r$  to the perturbation approaches the system  size $L$, we observe a saturation of the load change. This value decays  however  with $L$ with  a power law, 
 establishing  the saturation  as a finite size effect.

With a finite probability the
  maximal transmitted power  $\Delta {\rm Max}(F_{ij})$  increases when 
   a transmission line is added to the electricity grid, a phenomenon known as 
Braess' paradoxon.  This effect becomes more pronounced 
 when    the nonlinear equations Eq. \ref{dc} derived from Kirchhoff's law are considered
  rather than linearised equations as in previous studies \cite{witthaut2013}. 
   Induced changes in the load distribution on  AC grids and in more realistic grid topologies will  be studied in future work. In particular we shall study the role of randomness in the arrangement of consumers and producers in view of degrading the overall performance of the grid.

\end{document}